# Elastic and Strain-Tunable Electronic and Optical Properties of La$_2$AlGaO$_6$ Hybrid Perovskite: A First-Principles Study

Chaithanya Purushottam Bhat, Jyoti Dagar, Ashwin K. Godbole, and *Debashis Bandyopadhyay

Department of Physics, Birla Institute of Technology and Science Pilani, Rajasthan – 333031, India

*E-mail: bandy@pilani.bits-pilani.ac.in; debashis.bandy@gmail.com

**Abstract**

Perovskite materials, known for their structural versatility and multifunctional properties, continue to draw interest for advanced electronic and optoelectronic applications. In this study, we investigate the elastic and strain-engineered mechanical, electronic, and optical properties of the orthorhombic La$_2$AlGaO$_6$ (LAGO) hybrid perovskite using first-principles quantum mechanical calculations based on density functional theory (DFT). Structural optimizations were performed using both the local density approximation (LDA) and the generalized gradient approximation (GGA). The mechanical stability of LAGO was confirmed through the Born-Huang criteria, and key elastic constants (C$_{11}$, C$_{12}$, C$_{33}$, C$_{44}$, and C$_{66}$) were evaluated. These constants were further used to derive mechanical parameters, including Young's modulus, bulk modulus, shear modulus, Poisson's ratio, Cauchy's pressure, and anisotropic factor, which offer insights into the material's ductility, hardness, and elastic anisotropy. Crucially, we explored the influence of biaxial strain on the electronic band structure, DOS/PDOS, and Fermi energy, revealing significant band gap modulation under compressive and tensile strain, and hence, varying the optical properties. The coupling between elastic response and electronic structure highlights LAGO's potential for tunable device applications, where mechanical stimuli can be employed to tailor its electronic functionality.

## 1. Introduction

The investigation of perovskite materials has gathered significant interest in recent years, mainly due to their exceptional structural versatility and promising functionalities across numerous applications [1-6]. For their easily tunable electronic and optical properties, perovskites have found extensive utility in optoelectronics, catalysis, energy storage, and conversion, and in many other systems. Features, such as customizable band structures, high charge-carrier mobility, superior defect tolerance, and strong light-harvesting capabilities, have propelled their use in advanced technologies like solar cells, batteries, and energy storage [7-12]. Furthermore, doping techniques applied to perovskite materials have appeared as transformative tools, enabling fine-tuning of luminescence, enhancing structural stability, and optimizing device performance metrics such as External Quantum Efficiency (EQE) in light-emitting diodes [13-15]. These modifications highlight the critical role of dopants in improving the efficiency and durability of perovskite-based devices [16-20].

$ABO_3$-type perovskites and their hybrid counterparts, such as $A_2BB'O_6$ structures, exhibit unique crystal symmetries and electronic configurations that make them particularly appealing for both central research and practical applications [21-24]. Recent studies have established the efficacy of anion-doped perovskite oxides as catalysts in various chemical reactions, where alterations to the lattice structure, metal-oxygen bond strength, and defect configurations significantly influence catalytic activity [25, 26]. Furthermore, investigations into the electronic properties of these materials have emphasized the importance of orbital hybridization between transition metal 3d and oxygen 2p states, which governs their electronic behavior and stability [27]. Recent research on their different elastic coefficients, lattice vibration, and thermodynamic properties has deepened our understanding of their mechanical robustness and structural adaptability [28]. The ability to tune the electronic properties of perovskite materials using external factors, such as strain, presents an exciting opportunity for the development of advanced electronic devices. To study such important materials, Density Functional Theory (DFT) has emerged as one of the most effective computational methods for understanding the electronic, mechanical, and optical properties of these materials [29-35].

Study on the mechanical, electronic, and optical properties of LAGO perovskite, with an emphasis on their implications for diverse applications. Analyzing these properties is vital for appraising the material's toughness, damage resistance, and suitability for industrial purposes [36]. Computational techniques, including finite element analysis (FEA), serve as powerful tools to simulate material behavior under various conditions, providing predictive insights into mechanical performance through parameters such as Young's modulus [37, 38].

Elastic constants are essential for deriving key material attributes like sound velocity, Debye temperature, and melting point, all of which are integral to both scientific research and technological innovations. Additionally, changes in elastic properties during phase transitions offer valuable information on complex phenomena in systems such as heavy fermion materials and superconductors [39, 40]. The full anisotropic elastic tensor enables precise predictions of a material's mechanical response, thereby supporting the development of tailored composite materials with desired stiffness and elasticity profiles [41, 42]. In geophysics, elastic constants are instrumental in interpreting seismic data by correlating them with acoustic velocity measurements [43, 44].

Hooke's law, which defines the proportionality between stress and strain, forms the theoretical foundation for understanding elastic behavior in materials. The elastic constants $(C_{ij})$ relate stress $(\sigma)$ and strain $(\epsilon_i)$ as $(\sigma_i C_{ij})$ [45, 46]. The number of independent elastic constants depends on the material's symmetry. For tetragonal symmetry, as seen in LAGO perovskite, the constants include $C_{11}$, $C_{12}$, $C_{33}$, $C_{44}$, and $C_{66}$, while cubic symmetry involves $C_{11}, C_{12}$, and $C_{44}$. These constants provide critical insights into the mechanical stability, ductility, anisotropy, and elasticity of the material, thereby informing its practical applicability across various fields. In this study, we also employ DFT to investigate the effects of strain on the mechanical, electronic, and optical properties of $La_2AlGaO_6$ (LAGO) perovskite, providing valuable insights into its potential for strain-engineered electronic devices.

## 2. Computations

We conducted density functional theory (DFT) calculations utilizing the Vienna Ab initio Simulation Package (VASP) [47-50]. The interactions between ion cores and valence electrons were described using the projector augmented wave (PAW) method [51]. All calculations employed the Perdew-Burke-Ernzerhof (PBE) exchange-correlation functional [52] and included van der Waals interactions through Grimme's D2 dispersion correction [53]. Convergence criteria were set at $10^{-6}$ eV for energy and 0.005 eV/Å for forces on each atom. The plane-wave basis set had an energy cut-off of 550 eV. We have used the LAGO supercell $La_8Al_4Ga_4O_{24}$ with an 8×8×8 k-point mesh for relaxation. For density of states and band structure calculations, we employed the Γ-centered k-path. The calculated band structures of $La_2AlGaO_6$ perovskite by using the PBE functional are plotted with high symmetry points following the K-path as given, Γ→M→Z→ Γ→Z→R→A→A→Z|X→R|M→A in the first Brillouin zone.

## 3. Results and Discussion

### 3.1. Elastic Properties of LAGO Perovskite

In this study, our primary objective is to investigate the mechanical properties of pure LAGO perovskite by calculating various elastic constants. These constants are essential for characterizing the material's mechanical behavior under different conditions. In the present framework for our work, it is significant to acknowledge the work of Naher et al. [54], who conducted an extensive study on the mechanical properties of $Mo_5PB_2$. Their research provides valuable insights into the methodologies used for analyzing elastic properties in complex materials, which serve as a foundation for our investigation into LAGO perovskite. Their detailed work on $Mo_5PB_2$ has helped inform our approach to understanding the mechanical characteristics of perovskites and the significance of elastic constants in determining material behavior.

While examining the mechanical properties of materials with tetragonal structures, the mechanical stability can be measured using the Born stability criteria, referred to as the Born and Huang criteria [55]. These principles provide a set of necessary conditions that must be met for a material to be mechanically stable under minor perturbations. The stability conditions are expressed through specific constraints, which are precisely represented in equation (1). These constraints involve the material's elastic constants and ensure that the material will maintain structural integrity when subjected to external stresses, helping to predict whether the material will undergo any unstable deformations or transitions.

$$\left.\begin{array}{c} C_{11} > 0, C_{33} > 0, C_{44} > 0, C_{66} > 0 \\ C_{11} - C_{12} > 0, (C_{11} + C_{13} - 2C_{13}) > 0 \\ 2(C_{11} + C_{12})C_{33} + 4C_{13} > 0 \end{array}\right\} \quad (1)$$

$where, C_{11} = 230.133, C_{12} = 94.528, C_{13} = 94.659, C_{33} = 226.821, C_{44} = 78.649, C_{66} = 101.128$

The elastic constants $C_{11}$ and $C_{33}$ Characterize the resistance to linear compressions along the [100] and [001] directions, respectively, in LAGO. A higher value of $C_{11}$ compared to $C_{33}$ indicates a stronger bonding strength/compressibility along the [100] direction relative to the [001] direction in LAGO. Regarding shear deformation resistance, represented by the elastic constant $C_{44}$, it is notably lower than both $C_{11}$ and $C_{33}$ in LAGO. This suggests that LAGO is more susceptible to deformation via shear stress compared to unidirectional stress along its principal crystallographic directions. Observing that $C_{11}$ exceeds $C_{33}$ indicates stronger bonding strength/compressibility along the [100] direction compared to the [001] direction in LAGO. The elastic constant $C_{44}$, which signifies resistance to shear deformation, is notably smaller in LAGO compared to $C_{11}$ and $C_{33}$. This indicates that LAGO exhibits a greater tendency to deform under shear stress than under uniaxial stress along its principal crystallographic axes. Additionally, the elastic constant $C_{66}$ represents the shear resistance of the (100) plane in the [110] direction. In LAGO, the value of $(C_{11} + C_{12})$ exceeding $C_{33}$ suggests that the bonding within the (001) plane is mechanically more rigid than along the c-axis. Consequently, the elastic tensile modulus is higher in the (001) plane than along the c-axis.

### 3.1.1 Dynamical Stability ($S$):

The dynamic stability of a crystal depends on its shear constant, S, which is determined using the expression:

$$S = \frac{C_{11} - C_{12}}{2} \quad (2)$$

The shear constant $S$, also known as the tetragonal shear modulus, measures the stiffness of a material and its ability to resist shear deformation under stress in the (110) plane along the [110] direction. The sign of $S$ It is critical for assessing stability: a positive value indicates dynamical stability, whereas a negative value points to instability. For LAGO, the computed value of S is 67.8025 GPa, suggesting that the material is dynamically stable.

### 3.1.2 Kleinman parameter ($\xi$):

The Kleinman parameter ($\xi$) [56], Another important elastic parameter acts as a measure of internal strain, providing insight into the relative positioning of cation and anion sublattices during volume-preserving distortions, especially in cases where crystal symmetry does not impose constraints on atomic positions. This parameter evaluates a material's capacity to withstand strains resulting from stretching and bending. This parameter ($\xi$) of a compound can be calculated using the following correlation:

$\zeta = \frac{C_{11} + 8C_{12}}{7C_{11} + 2C_{12}} - - - -(3)$

The Kleinman parameter ($\xi$) spans from zero to one ($1 \leq \xi \leq 1$) for a solid $\xi$ values of 0 and 1 denote the lower and upper boundaries, respectively. A $\xi$ value of 0 signifies minimal bond bending contribution to endure external stress, whereas $a \, \xi$ equal to 1 indicates minimal bond stretching/contracting contribution to resist externally applied stress. The computed $\xi$ for LAGO is 0.55, indicating that the mechanical strength of LAGO is influenced by both bond bending and bond stretching/contracting contributions.

### 3.1.3 Bulk modulus ($B$) and Shear modulus ($G$):

The bulk modulus ($B$) measures a material's resistance to changes in volume when subjected to applied pressure, while the shear modulus ($G$) evaluates its resistance to reversible deformations resulting from shear stress. As a result, ($G$) serves as a more accurate indicator of a material's hardness than ($B$). A higher value of the shear modulus suggests stronger directional bonding between atoms. In the case of a crystalline aggregate made up of single-phase crystals with random orientations, the distribution of stress or strain under external forces can be analyzed using two extreme approaches. One approach, known as the Voigt approximation, assumes uniform strain across the polycrystalline aggregate in response to external strain. The other, called the Reuss approximation, assumes uniform stress across the material when subjected to external stress.

The general expressions for Voigt bulk modulus ($B_V$) and shear modulus ($G_V$), and Reuss bulk modulus ($B_R$) and shear modulus ($G_R$) are as follows:

$$B_V = \frac{1}{9}(C_{11} + C_{22} + C_{33}) + \frac{2}{9}(C_{12} + C_{13} + C_{23}) \qquad (4)$$

$$G_V = \frac{1}{15}(C_{11} + C_{22} + C_{33}) + \frac{1}{5}(C_{44} + C_{55} + C_{66}) - \frac{1}{15}(C_{12} + C_{13} + C_{23}) \qquad (5)$$

$$\frac{1}{B_R} = (C_{11} + C_{22} + C_{33}) + 2(C_{12} + C_{13} + C_{23}) \qquad (6)$$

$$\frac{1}{G_R} = \frac{4}{15}(C_{11} + C_{22} + C_{33}) + \frac{1}{5}(C_{44} + C_{55} + C_{66}) - \frac{4}{15}(C_{12} + C_{13} + C_{23}) \qquad (7)$$

Hill illustrated that the Voigt and Reuss equations act as upper and lower limits for the actual polycrystalline constants. He proposed that a pragmatic estimation of the bulk and shear moduli could be attained by averaging these extremes. Hence, the elastic moduli can be approximated using Hill's average as follows:

$$B_H = \frac{1}{2}(B_V + B_R) \qquad (8); \; G_H = \frac{1}{2}(G_V + G_R) \qquad (9)$$

Different calculated parameters are given in the following Table 1.

| Mechanical Properties | Voigt | Reuss | Hill |
|---|---|---|---|
| Bulk Modulus: $B$ (GPa) | 139.42 | 139.414 | 139.417 |
| Young's Modulus: $E$ (GPa) | 198.43 | 194.718 | 196.577 |
| Shear Modulus: $G$ (GPa) | 78.57 | 76.829 | 77.698 |

By using these parameters, other elastic parameters can be defined, measured, and discussed in the following sections.

### 3.1.4 Young's modulus ($E$):

Young's modulus is a key parameter for assessing the stiffness of a solid material, and it can be determined using the following equation:

$$E_X = \frac{9B_X G_X}{3B_X + G_X} \qquad (10)$$

Where subscripts $X = V; \; R; \; H$ represent Voigt, Reuss, and Hill approximations, respectively.

### 3.1.5 Poisson's ratio ($v$):

Poisson's ratio ($v$), as expressed in equation (11), is an essential parameter for evaluating a material's compressibility, brittleness, or ductility, and bonding characteristics. This ratio reflects how a material deforms in the perpendicular direction when subjected to stress. In terms of brittleness and ductility, a material is generally considered brittle if its Poisson's ratio is less than or equal to 0.33, whereas a Poisson's ratio above this threshold typically indicates ductility. For the compound LAGO, with a calculated Poisson's ratio of 0.26, the material exhibits brittle behavior. Additionally, Poisson's ratio provides valuable insight into the nature of bonding within a material. Ionic and covalent materials tend to have distinct values for Poisson's ratio, with typical values around 0.25 for ionic materials and 0.10 for covalent materials. The Poisson's ratio of LAGO, being closer to the value for ionic materials, suggests a significant ionic component in its chemical bonding. This observation highlights the influence of ionic interactions in the structural characteristics of LAGO, further contributing to its mechanical properties.

$$v_X = \frac{3B_X - 2G_X}{2(3B_X + G_X)} \qquad (11)$$

### 3.1.6 Pugh's ratio $\left(\frac{B}{G}\right)$:

Once more, Pugh introduced a simple correlation to predict whether materials exhibit brittle or ductile characteristics, proposing that the ratio of bulk to shear modulus $\left(\frac{B}{G}\right)$ serves as an indicator. Pugh established a critical threshold for the transition from ductile to brittle behavior: if $\frac{B}{G} \leq 1.75$, the material behaves brittlely; otherwise, it behaves ductilely. For LAGO, with a $\frac{G}{B}$ value of 1.77, slightly exceeding the specified condition, we anticipate the compound to demonstrate brittle behavior.

### 3.1.7 Cauchy pressure $(C^{\circ})$:

The Cauchy pressure $(C^{\circ})$ of a material is expressed as

$$C^{\circ} = C_{12} - C_{44} \qquad (12)$$

A positive Cauchy pressure usually indicates ductility, whereas a negative value suggests brittleness. In the instance of LAGO, the positive Cauchy pressure strongly indicates the compound's brittleness. Furthermore, Cauchy pressure is employed to describe the atomic bonding along the angular direction in a material. If this value is positive, it indicates ionic bonding, and for a negative value, it is covalent bonding in nature. In LAGO, with a Cauchy pressure value of 15.9 GPa, it exhibits bonding similar to ionic bonding.

### 3.1.8 Machinability index $(\mu_M)$:

The machinability index also provides insights into a material's plasticity and dry lubrication behavior. The equation suggests that a combination of high overall strength and low shear resistance enhances machinability and dry lubrication performance. Materials with a high $\mu_M$ value exhibit superior dry lubrication, reduced feed forces, lower friction coefficients, and improved plastic deformation potential. With a $\mu_M$ value of 1.77, LAGO is predicted to exhibit excellent machinability, comparable to that of technologically significant MAX phase compounds, alloys, and other layered ternaries.

$$\mu_M = \frac{B}{C_{44}} \qquad (13)$$

The machinability index also provides insights into a material's plasticity and dry lubrication behavior. The equation suggests that a combination of high overall strength and low shear resistance enhances machinability and dry lubrication performance. Materials with a high $\mu_M$ value exhibit superior dry lubrication, reduced feed forces, lower friction coefficients, and improved plastic deformation potential. With a $\mu_M$ value of 1.77, LAGO is predicted to exhibit excellent machinability, comparable to that of technologically significant MAX phase compounds, alloys, and other layered ternaries.

**893.1.9 Debye temperature and Sound velocity ($\theta_D$):**

The Debye temperature is a crucial physical property that affects various characteristics of solid materials, such as the melting point, thermal conductivity, specific heat, vacancy formation energy in metals, and superconducting transition temperature. It can be calculated using the following equation:

$$\theta_D = \frac{h}{k_B} \left[ \left( \frac{3n}{4\pi} \right) \frac{N_A \rho}{M} \right]^{\frac{1}{3}} v_m \qquad (14)$$

Where $h$ is Planck's constant, $k_B$ is the Boltzmann constant, $n$ represents the number of atoms in the molecule, $N_A$ is Avogadro's number, $\rho$ is the density, and $M$ is the molar mass.

The average velocity of the sound wave $v_m$, can be determined as:

$$v_m = \left[ \frac{1}{3} \left( \frac{2}{v_t^3} + \frac{1}{v_l^3} \right) \right]^{-\frac{1}{3}} v_m \qquad (15)$$

Where the transverse ($v_t$) and longitudinal ($v_l$) Sound velocities are obtained using the shear modulus $G$ and the bulk modulus $B$, which can be calculated as follows:

$$v_t = \sqrt{\frac{G}{\rho}} \qquad (16); v_l = \sqrt{\frac{3B + 4G}{3\rho}} \qquad (17)$$

The calculated values are listed in Table 2.

| | $C''$ (GPa) | $\mu_M$ | $\Theta_D$ (K) | $v_m$ (m/s) | $v_t$ (m/s) | $v_l$ (m/s) |
|---|---|---|---|---|---|---|
| LAGO | 15.9 | 1.77 | 502.4 | 3907.037 | 3513.001 | 6212.822 |

**3.1.10 Melting temperature $T_m$:**

The melting temperature of a crystalline solid is closely related to its thermal expansion and bonding energy. Typically, materials with higher melting points exhibit stronger atomic interactions, higher bonding and cohesive energies, with lower thermal expansion coefficients. Furthermore, the melting temperature provides an initial estimate of the maximum operational temperature for materials before significant oxidation, chemical alterations, or excessive deformation occur. The melting temperature can be estimated using the elastic constants as follows:

$$T_m = \alpha \sqrt{G.B} \qquad (18)$$

Where $\alpha$ is a proportionality constant, $G$ is the shear modulus, and $B$ is the bulk modulus. With an approximate melting temperature of 1384.63±300K for LAGO, the material shows promise for high-temperature applications. Nevertheless, it's crucial to acknowledge that this estimated melting temperature should be treated as a rough approximation due to significant uncertainty, particularly given the lack of experimental values for comparison at present.

### 3.1.11 Elastic anisotropy:

The universal anisotropy factor has become increasingly popular in recent years as an index for anisotropy, owing to its simplicity compared to the various specific anisotropy factors designated for individual crystal planes. Ranganathan and Ostoja-Starzewski [57] proposed the notion of the universal anisotropy index, represented as $A^U$, which provides a unified measure of anisotropy irrespective of crystal symmetry. It is determined by the equation:

$$A^U = 5\frac{G_V}{G_R} + \frac{B_V}{B_R} - 6 \geq 0 \qquad (19)$$

To derive a singular anisotropy measure, Chung and Buessem **[58]** introduced the following empirical approach,

$$A^C = \frac{G_V - G_R}{G_V + G_R} \qquad (20)$$

The universal anisotropy factor is limited to values of zero or greater. A value of zero for $A^U$ denotes isotropy within the crystal, whereas any non-zero value indicates the existence of anisotropy. Usually, in a low-range isotropic crystal, $A^U$ equals $A^C$, and both are zero. In the case of LAGO, where $A^C$ is 0.0 and $A^U$ is 0.11, this suggests that the material exhibits minimal anisotropy.

## 3.2. Effect of strain on electronic properties

### 3.2.1. Strain-Dependent Electronic Structure of La$_2$AlGaO$_6$

To clarify the effect of uniaxial strain on the electronic properties of La$_2$AlGaO$_6$, we methodically investigated its density of states, band structure, and related electronic parameters under varying strain magnitudes ranging from -12% to +12%, applied along the crystallographic c-axis. The results disclose noticeable alterations in the electronic behavior, particularly in the band gap, Fermi energy, frontier orbital positions, and hybridization strength. Optimized supercell and corresponding charge distributions of LAGO under the no-strain condition are shown in Figure 1 below.

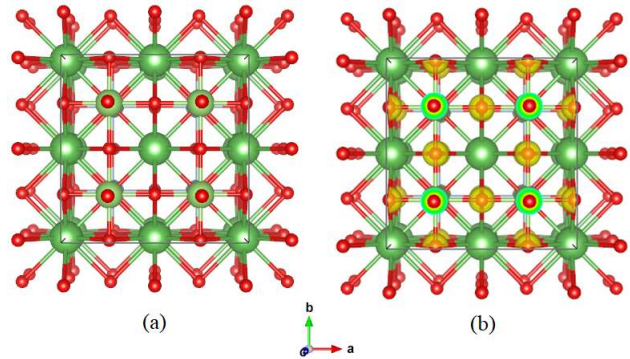

Figure 1. Structure of La$_2$AlGaO$_6$ along with its (b) charge distributions.

These changes are strongly dependent on both the magnitude and sign of the applied strain, highlighting the asymmetric response of the system to compressive and tensile deformation. The strain-dependent band structure of LAGO perovskite is shown in Figures 2a and 2b.

For clarity and comprehensive understanding, the discussion of the strain-dependent electronic structure is organized into four subsections. Each subsection corresponds to specific strain regimes and includes a comparative analysis between equivalent compressive and tensile strains (e.g., ±4% and ±6%). Particular emphasis is placed on the evolution of orbital hybridization, shifts in the valence and conduction band edges, symmetry-driven transitions, and the structural origins of observed discontinuities in electronic parameters. This detailed breakdown allows for an in-depth interpretation of the strain-induced modulation of the electronic landscape in $La_2AlGaO_6$, paving the way for its potential application in strain-tunable electronic and optoelectronic devices.

### 3.2.2. Strain-Dependent Electronic Structure and Frontier Orbital Hybridization

The effect of uniaxial strain applied along the c-axis (z-direction) on the electronic structure of $La_2AlGaO_6$ was systematically studied across a wide strain range from –12% to +12%. The strain-induced modifications in band structures (Figures 2), as well as the variations in band gap, Fermi energy, total energy, binding energy, and frontier band edge positions (Figure 3), provide crucial insights into the electronic behavior of the system under elastic deformation. Notably, the observed trends are highly asymmetric concerning compressive and tensile strain, emphasizing the anisotropic nature of the lattice response and the electronic configuration of the constituent atoms.

### 3.2.2.1. Electronic Evolution Under Mild Strain (0 to –4% and 0 to +4%)

In the strain interval between -4% to +4%, the electronic structure of $La_2AlGaO_6$ exhibits quasi-linear and reversible behavior. The calculated band structures remain semiconducting, with no apparent closure of the band gap or abrupt changes in orbital character. The band gap varies moderately, from approximately 3.10 eV at –4% compressive strain to 2.90 eV at +4% tensile strain. This slight asymmetry in band gap evolution reflects the intrinsic non-centrosymmetric bonding environment in the perovskite lattice. Under mild compressive strain (–4%), the decrease in lattice parameter along the z-axis results in shortened La–O, Al–O, and Ga–O bond lengths. This compression enhances the crystal field splitting and the orbital hybridization between O 2p states and the unoccupied cationic orbitals (primarily La 5d, Ga 4s, and Al 3s). The increased orbital overlap leads to a greater energy separation between the bonding and antibonding states, resulting in an upward shift of the conduction band minimum (CBM) and a corresponding increase in band gap.

Conversely, under an equivalent tensile strain (+4%), bond elongation leads to a decrease in p–d and p–s hybridization, weakening the electronic coupling between the anions and cations. This causes a reduction in crystal field strength and narrows the band gap due to the downward movement of the CBM. However, the VBM position remains relatively stable, as it is dominated by strongly localized O 2p states. Overall, this strain range represents a nearly elastic regime where the system maintains its structural symmetry, and the changes in electronic properties are driven primarily by bond length modulation.

### 3.2.2.2. Asymmetric Behavior Under Moderate Strain (-4% to −6% and 4% to +6%)

The application of moderate strain magnitudes ($|\varepsilon| = 6\%$) leads to a nonlinear and asymmetric evolution of the electronic structure, reflecting the onset of anisotropic lattice distortion and partial breakdown of orbital degeneracy. For 6% comprehensive strain, the band gap reaches its maximum value (~3.35 eV), accompanied by the highest Fermi energy observed in the compressive regime (~7.5 eV). The band structure becomes more dispersive, especially around the CBM, indicating an enhancement in carrier mobility. This behavior is consistent with a further strengthening of orbital hybridization under compression, particularly involving the antibonding interactions between O 2p and Ga/Al s-d states. The enhanced orbital repulsion pushes the CBM higher, increasing the band gap. In contrast, at +6% tensile strain, the band gap drops to ~2.75 eV, and the Fermi level shifts downward to ~6.1 eV. This reduction is associated with a reduction in orbital overlap due to increased interatomic distances, leading to a lower CBM energy and flatter conduction band, which indicates reduced carrier delocalization and higher effective masses.

A direct comparison of +6% and -6% strain reveals a clear electronic asymmetry, with compressive strain inducing a more significant band gap widening than the narrowing induced by the same magnitude of tensile strain. This asymmetric response can be attributed to the non-linear coupling between strain and internal octahedral distortions, which affects orbital orientations and interactions differently in tension and compression.

### 3.2.3. Role of Hybridization and Frontier Orbital Dynamics

The observed electronic modulation under strain is intrinsically governed by the hybridization dynamics of frontier orbitals. The valence band maximum is predominantly composed of nonbonding O 2p orbitals with minor mixing from Ga 4p and Al 3p states, while the conduction band minimum originates from La 5d, Ga 4s, and Al 3s states.

Under compressive strain, the contraction of octahedra enhances the orbital overlap, increasing the crystal field strength and promoting stronger bonding-antibonding splitting. This leads to the lifting of CBM and deepening of VBM, enhancing the electronic band gap and stabilizing the semiconducting state. The bands near the Fermi level are more dispersive in this regime, suggesting increased orbital delocalization and higher carrier mobility. In contrast, tensile strain reduces the hybridization strength as the metal–oxygen bond lengths elongate, leading to a lower CBM and a mild upward shift in the VBM. The frontier orbitals become less hybridized, resulting in more localized electronic states. The flattening of bands near the Fermi level-most prominent beyond +6% - is indicative of suppressed electronic coupling and may signal a transition toward semimetallic or low-mobility electronic behavior.

This evolution in orbital interaction strength is reflected in the trends of the conduction and valence band edge positions (Figure 3, bottom panel), where the CBM drops significantly more under tensile strain than the VBM rises, accounting for the net band gap reduction.

### 3.2.4. Strain-Induced Electronic Properties and Orbital Hybridization

The strain-dependent TDOS and PDOS of $La_2AlGaO_6$ perovskite, subjected to uniaxial strain from –8% to +8% along the c-axis, exhibit pronounced variations in the electronic structure, driven primarily by modifications in the orbital hybridization between constituent atoms (Figure 3). At –8% compressive strain, a noticeable narrowing of the band gap is observed. This behavior arises from enhanced orbital overlap due to reduced interatomic distances, particularly between the O-2p orbitals in the valence band and the Al-3p and Ga-4p orbitals in the conduction band. The PDOS clearly shows that the O-2p states dominate the upper valence band, while the conduction band is composed of significant contributions from Ga-4s, Ga-4p, and Al-3s, with minor contributions from La-d states. Under compression, the increased overlap between these orbitals leads to stronger hybridization, broadening the band edges and resulting in increased electronic delocalization. This indicates a shift toward a more covalent bonding character, which is reflected in the enhanced density of states near the Fermi level.

As the strain transitions from compressive to tensile, especially from 0% to +8%, the hybridization behavior shifts significantly. Tensile strain elongates the lattice along the c-axis, reducing orbital overlap and thus weakening the hybridization between O-2p and Al/Ga p-orbitals. Consequently, the conduction band edge moves to higher energy, and the valence band becomes more localized, resulting in a wider band gap. The PDOS under tensile strain shows reduced intensity and sharper peaks, indicating more localized electronic states due to diminished orbital mixing. In particular, the weakening of Al–O and Ga–O bonds under tension suppresses the p–p and p–s interactions responsible for the original band structure. The gradual decoupling of these orbitals leads to a more ionic character and reduced carrier mobility. Most significantly, across the entire strain range, the La-f states (omitted in the right column of the PDOS for clarity) are not significantly involved in the states near the band edges and thus do not influence the electronic transitions directly. The electronic behavior is predominantly governed by the hybridization between the oxygen 2p orbitals and the p/s orbitals of Al and Ga. The ability to modulate this hybridization through strain engineering highlights the material's potential for electronic and optoelectronic device applications, such as strain-sensitive transistors or band gap-engineered photovoltaics.

### 3.2.4. Discontinuities in Parameter Evolution: Structural Origins and Consequences

Despite the overall continuity of strain-induced trends, several discontinuities or abrupt changes in band gap, Fermi level, and binding energy are observed, particularly around ±6% and ±10%. For instance, the band gap shows an unexpected dip at +6%, and the Fermi energy exhibits a sudden decline between +6% and +8%, deviating from the otherwise smooth trend.

Similarly, the binding energy curve (Figure 3, middle panel) features noticeable inflection points that break the expected parabolic symmetry. These discontinuities likely originate from nonlinear structural distortions or octahedral reconstructions induced by critical strain thresholds. At specific strain values, the rigid $BO_6$ octahedra (B = Al, Ga) undergo tilting, rotation, or buckling, which can break local symmetry and lead to reorganization of the crystal field environment. These symmetry breakings modulate the electronic band structure in a non-continuous manner by altering orbital overlap conditions, hybridization strengths, and energy level orderings.

Moreover, the transition from a highly hybridized to a weakly hybridized regime may cause certain electronic states to shift rapidly in energy, manifesting as band reordering or band crossing phenomena. These structural effects are more pronounced under tensile strain due to the softening of the lattice and reduction in octahedral rigidity. Hence, the discontinuous trends in electronic parameters serve as signatures of incipient phase transitions or electronic reconstructions driven by strain-induced lattice instabilities, underscoring the need for careful strain control in practical applications.

This comprehensive analysis reveals that $La_2AlGaO_6$ is an electronically versatile material, where strain acts as a tuning knob for band gap, band edge positions, Fermi level, and hybridization. The asymmetric and nonlinear response to tensile and compressive deformation, coupled with hybridization-driven band structure evolution, offers a rich platform for strain-engineered optoelectronic applications, such as tunable insulators, photodetectors, or piezoelectric devices.

**Table 1: Strain-dependent Band gap (Gap), Fermi Energy ($E_f$), $E-E_0$, and Binding Energy (BE). Strain is applied along the C-axis (Vertical) of the Materials**

| Strain | c (Å) | Gap (eV) | $E_f$ (eV) | $E-E_0$ (eV) | BE (eV) | Strain | c (Å) | Gap (eV) | $E_f$ (eV) | $E-E_0$ (eV) | BE (eV) |
|---|---|---|---|---|---|---|---|---|---|---|---|
| 0 | 8.353 | 3.267 | 6.276 | 0 | 1.255 | 0 | 8.353 | 3.267 | 6.276 | 0 | 1.255 |
| -1 | 8.868 | 3.256 | 6.396 | 0.315 | 0.94 | 1 | 7.839 | 3.341 | 6.166 | -0.24 | 1.496 |
| -2 | 8.71 | 3.135 | 6.516 | 0.527 | 0.728 | 2 | 7.76 | 3.052 | 6.156 | -0.816 | 2.072 |
| -3 | 8.551 | 3.065 | 6.636 | 0.542 | 0.713 | 3 | 7.681 | 3.162 | 5.944 | -0.966 | 2.222 |
| -4 | 8.393 | 2.994 | 6.766 | 0.689 | 0.566 | 4 | 7.601 | 2.803 | 6.086 | -1.876 | 3.132 |
| -6 | 8.235 | 2.842 | 7.026 | 0.395 | 0.86 | 6 | 7.443 | 2.561 | 6.016 | -3.125 | 4.38 |
| -8 | 8.155 | 2.688 | 7.277 | -0.434 | 1.69 | 8 | 7.284 | 2.242 | 5.956 | -4.516 | 5.772 |
| -10 | 8.076 | 2.527 | 7.539 | -2.007 | 3.263 | 10 | 7.126 | 2.607 | 5.158 | -2.94 | 4.195 |
| -12 | 7.997 | 2.364 | 7.807 | -3.504 | 4.76 | 12 | 6.968 | 2.532 | 4.946 | -3.31 | 4.566 |

### 3.1.4. Implications of Strain-Modulated Frontier Orbitals

The modulation of the HOMO and LUMO energy levels under strain reveals a strong coupling between the lattice and the electronic properties of $La_2AlGaO_6$. This relationship is governed by how strain alters interatomic distances, orbital overlaps, and the degree of hybridization between constituent atomic orbitals, ultimately controlling the material's band gap and electronic dispersion. Under compressive strain, the reduction in volume and hence in the bond lengths, especially within the $GaO_6$ and $AlO_6$ octahedra, enhances the overlap between oxygen 2p orbitals and cation states such as Ga-4s, Al-3p, and La-5d. This increased hybridization leads to stronger orbital coupling and electronic delocalization, particularly at the band edge. The resulting downward shift in the LUMO energy narrows the band gap, which benefits applications that require efficient charge transport. Additionally, the improved electronic connectivity supports lower activation energies for charge carriers, enhancing electrical conductivity. However, tensile strain stretches the lattice and increases bond lengths, which weakens orbital overlap and reduces the extent of hybridization. This leads to more localized electronic states, particularly at the valence band edge (HOMO), which shifts to lower energies. Simultaneously, the diminished interaction between La-5d and Ga-4s orbitals causes the LUMO to shift upward in energy, resulting in a wider band gap. These strain-dependent changes in electronic structure illustrate the versatility of $La_2AlGaO_6$ as a strain-engineerable perovskite. By tuning the biaxial strain, through epitaxial growth on mismatched substrates or external mechanical stress, researchers can systematically manipulate the positions of the HOMO and LUMO to tailor the band alignment and carrier dynamics. This level of control provides a powerful strategy for optimizing device performance across a wide range of applications, from high-efficiency solar cells to UV detectors and dielectric materials. The above variation can be understood with the help of the DOS/PDOS variation of LAGO with the applied strain, as shown in Figure 4. The contribution of La-f is not shown in the figure because of its dominating contributions. With the application of the strain, the HOMO-LUMO gaps have been changed because of the variation of HOMO, which contains mainly Ga-d contributions, as shown in the DOS plot.

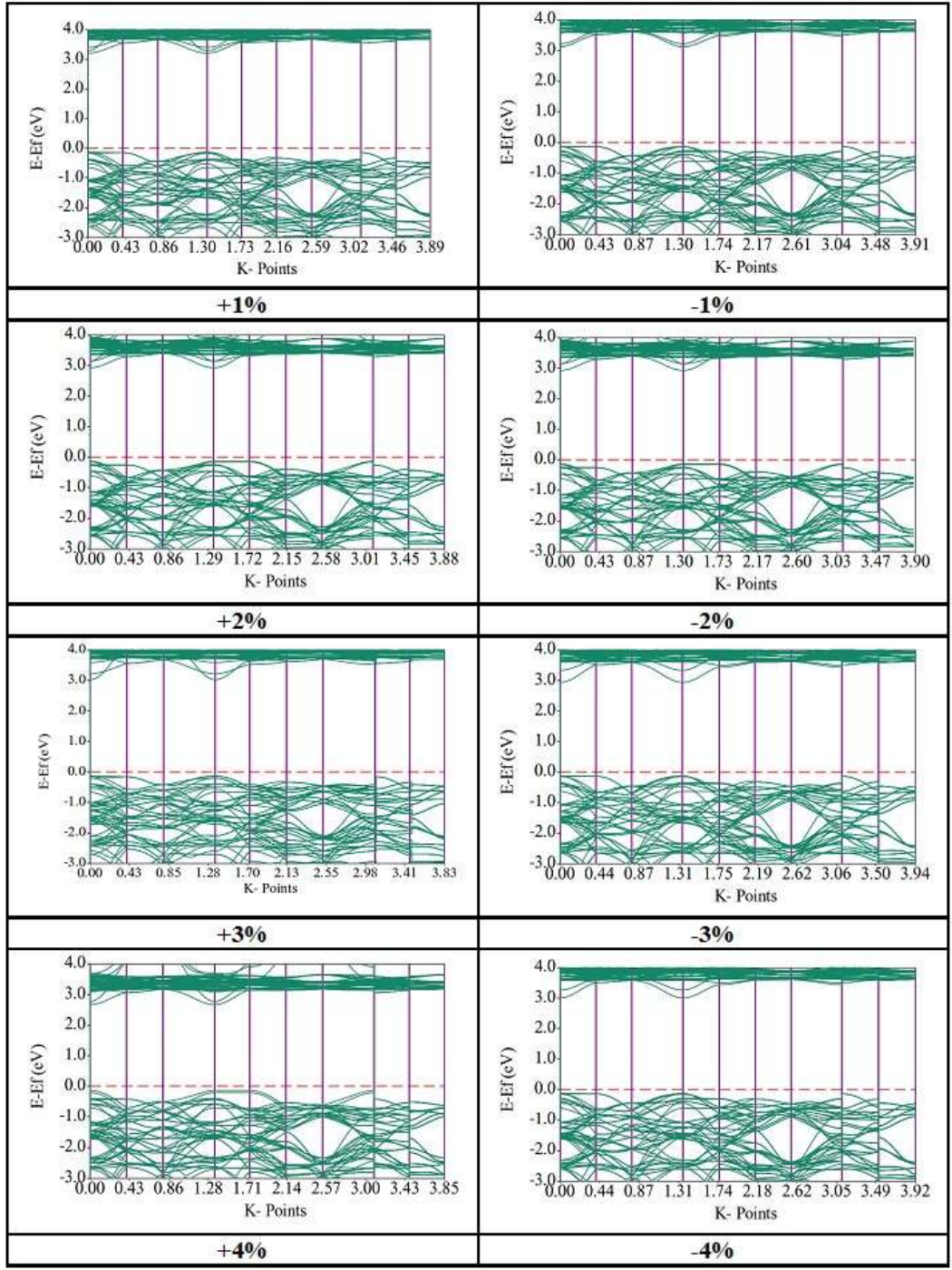

**Figure 2a. Band structures of LAGO under different applied strains along the c-axis with the K path Γ→M→Z→Γ→Z→R→A→Z|X→R|M→A**

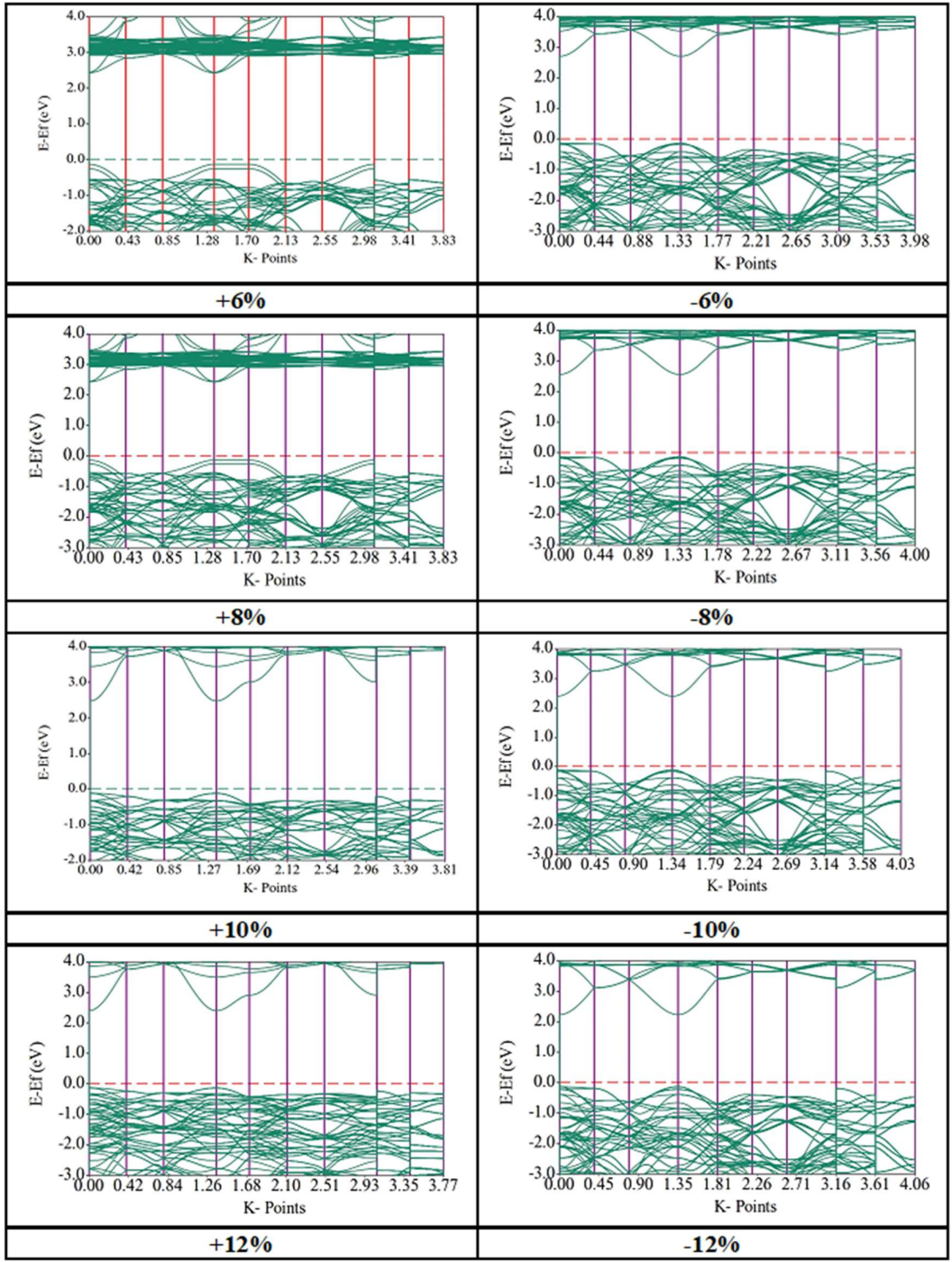

**Figure 2b. Band structures of LAGO under different applied strains along the c-axis with the K path Γ→M→Z→Γ→Z→R→A→Z|X→R|M→A**

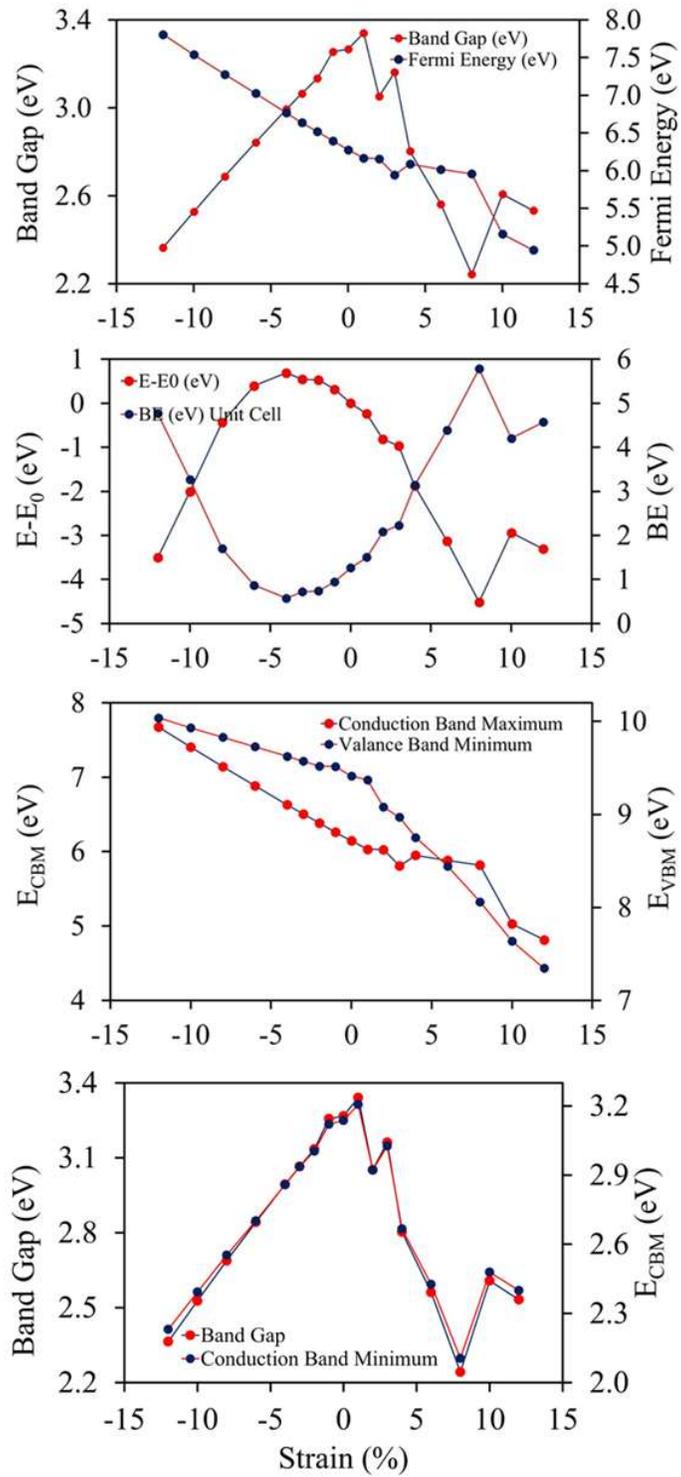

**Figure 3. Variation of different parameters under applied strain**

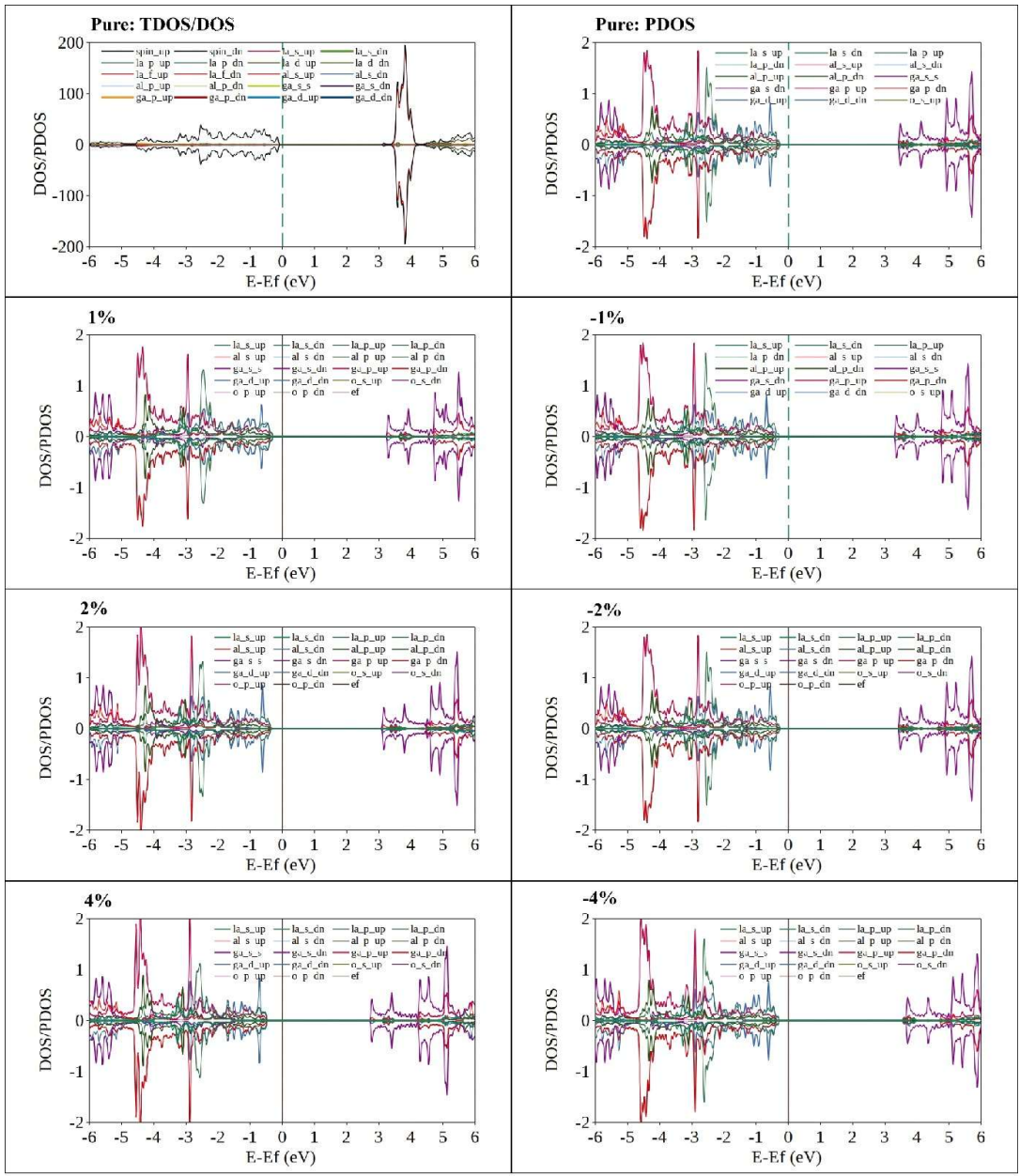

**Figure 4.** Strain-induced DOS Plot with the variation of applied strain. Due to the dominating contribution of La-f, the figures are plotted without La-f contributions.

**Figure 4. Continued.**

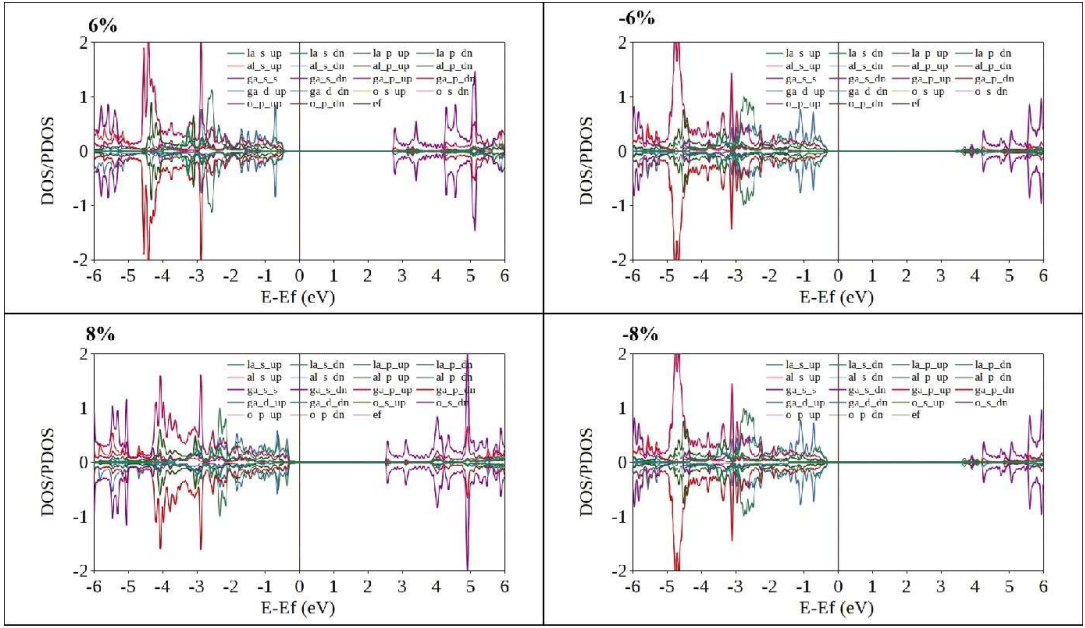

## 3.4 Strain-dependent optical properties of LAGO (uniaxial strain along z, −8% → +8%)

To investigate the influence of mechanical deformation on the optical response of La₂AlGaO₆ (LAGO) perovskite, we analyzed its optical properties under uniaxial strain ranging from −8% (compressive) to +8% (tensile) along the crystallographic c-axis. The variation of different optical parameters is shown in Figure 4. The optical parameters were computed using the frequency-dependent complex dielectric function, from which the real and imaginary parts ($\varepsilon_1(\omega)$ and $\varepsilon_2(\omega)$), as well as derived properties, such as the refractive index, absorption coefficient, reflectivity, optical conductivity, and energy loss function, were evaluated. The relation between different optical parameters and the real and imaginary parts of the dielectric constants is shown in Table 2.

**Absorption ($\alpha(\omega)$):** Under compressive strain (toward −8%), the fundamental absorption edge moves to higher photon energies (blue-shift) and the UV absorption peaks sharpen and increase in intensity. This indicates that compressive deformation enhances specific interband transition matrix elements and increases oscillator strength for O-2p → La/Al/Ga conduction-band transitions in the UV. In the visible range, the material remains essentially transparent for all strains, but compressive strain produces a measurable increase in near-edge UV absorption. As strain is relaxed toward 0% and then into tensile (+) values, the absorption edge shifts toward lower photon energies (red-shift) and peak intensities decrease. At large tensile strain (+8%), the absorption onset is at the lowest photon energies across the series, and absorption peaks are broadened and weakened-consistent with reduced transition probability and modified band dispersion. The net effect across −8 → +8: compressive → stronger, higher-energy UV absorption; tensile → weaker, lower-energy (red-shifted) UV absorption, and increased transparency.

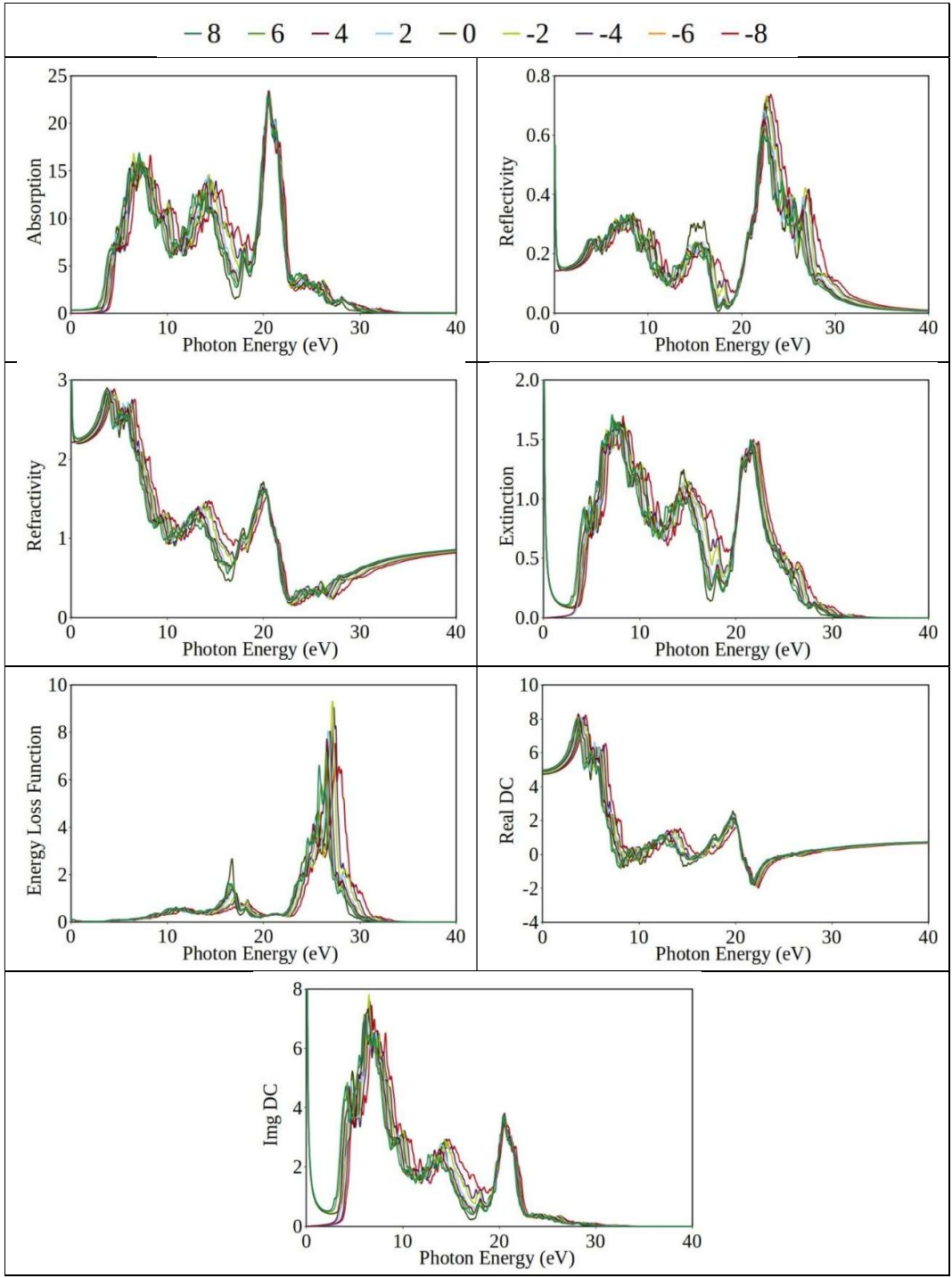

**Figure 4.** Variation of different optical parameters with the induced positive and negative strain along the z-axis (vertical axis). The strain-dependent color code for the graphs is shown at the top of the figure.

**Table 2:** relation between different optical parameters with real and imaginary parts of dielectric constants.

| Parameter | Expression | Physical Meaning | Parameter | Expression | Physical Meaning |
|---|---|---|---|---|---|
| Refractive index (n) | $\left[\dfrac{\sqrt{\varepsilon_1^2 + \varepsilon_2^2}+\varepsilon_1}{2}\right]^{\frac{1}{2}}$ | Phase velocity | Reflectivity (R) | $\dfrac{(n-1)^2 + \kappa^2}{(n+1)^2 + \kappa^2}$ | Surface reflection |
| Extinction coefficient ($\kappa$) | $\left[\dfrac{\sqrt{\varepsilon_1^2 + \varepsilon_2^2}-\varepsilon_1}{2}\right]^{\frac{1}{2}}$ | Absorption strength | Absorption coefficient ($\alpha$) | $\dfrac{4\pi}{\lambda}\left[\dfrac{\sqrt{\varepsilon_1^2 + \varepsilon_2^2}-\varepsilon_1}{2}\right]^{\frac{1}{2}}$ | Light attenuation |
| Dielectric Function $\epsilon(\omega)$ | $\sqrt{\varepsilon_1^2 + \varepsilon_2^2}$ | Dielectric Constant | | | |

**Reflectivity (R($\omega$)):** Compressive strain increases low-energy reflectivity: for negative strains, the reflectivity in the near-UV/visible border rises noticeably (a stronger reflectivity peak appears near the absorption edge). This follows from larger oscillator strengths and enhanced dielectric response under compression. Tensile strain suppresses reflectivity across the visible and near-UV: R($\omega$) drops, and the reflectivity peak near the band edge becomes less pronounced, producing a more transparent surface (beneficial for antireflective/transparent coatings). Across the full $-8 \rightarrow +8$ sweep, reflectivity shows a monotonic decrease with increasing tensile strain.

**Refractive index (n($\omega$)):** The static (low-energy) refractive index increases slightly under compressive strain, reflecting enhanced polarizability and stronger dispersion below the absorption edge. Thus, at $-8\%$ the peak/low-energy n($\omega$) is highest. As strain moves toward tensile ($+8\%$), the refractive index in the visible range decreases gradually (normal dispersion behavior maintained) and the sharp structure in n($\omega$) near the absorption edge shifts to lower energy (consistent with the red shift of the absorption edge). In short: compression $\rightarrow$ modest increase in n (stronger dispersion); tension $\rightarrow$ reduced n (more optically transparent).

**Real and imaginary parts of the dielectric function ($\varepsilon_1(\omega)$, $\varepsilon_2(\omega)$):** $\varepsilon_1(\omega)$: The static dielectric constant ($\varepsilon_1$ at $\omega \rightarrow 0$) rises under compressive strain, indicating enhanced dielectric screening. The energy at which $\varepsilon_1(\omega)$ crosses zero (signaling the onset of metallic-like behavior/plasma resonance) shifts to higher energies under compression and to lower energies under tension. Thus, compressive strain pushes the dielectric $\rightarrow$ metal crossover to higher photon energies. $E_2(\omega)$: The imaginary part (absorption-related) shows distinct UV peaks corresponding to O-2p $\rightarrow$ La/Al/Ga transitions. Under compression, these $\varepsilon_2$ peaks become stronger and shift to higher energies (blue-shift); under tension, they weaken and shift to lower energies (red-shift). Peak widths also evolve: compressive strain typically sharpens dominant $\varepsilon_2$ peaks (indicating more well-defined interband transitions), while tensile strain broadens them (reduced joint-density-of-states coherence).

**Extinction coefficient (k($\omega$)):** At low photon energies, k($\omega$) remains near zero for all strains (visible transparency retained). Near the absorption edge, k rises abruptly: compressive strain produces larger peak magnitudes and blue-shifted peak positions, whereas tensile strain reduces peak heights and red-shifts them. Across $-8 \rightarrow +8$, the extinction behavior is therefore a direct mirror of $\varepsilon_2$ trends: more absorption strength and higher-energy peaks under compression; weaker, lower-energy absorption under tension.

**Energy loss function ($L(\omega) = -\text{Im}(1/\varepsilon(\omega))$) and Plasmonic behavior:** The bulk plasmon (dominant $L(\omega)$ peak) is strain-sensitive. Under compressive strain ($-8\%$), the main plasmonic peak shifts to higher photon energies and often becomes more pronounced, reflecting increased effective electron density and stronger collective oscillation energy. Under tensile strain ($+8\%$), the plasmon peak shifts toward lower photon energies and its intensity is reduced (plasmon damping increases). Secondary shoulders or low-energy features that arise from interband contributions also shift consistently (blue under compression, red under tension). Net effect: compression moves the plasmon resonance up in energy and sharpens it; tension pushes it down and suppresses it.

Combined trends, physical picture, and applications: Compressively strained LAGO (negative z-strain, approaching $-8\%$) behaves optically as a more strongly interacting, more strongly screening dielectric with enhanced UV absorption, higher refractive index, larger $\varepsilon_1(0)$, stronger extinction in the UV, elevated/plasmon energies, and increased low-energy reflectivity. These attributes favor optoelectronic and photocatalytic uses where stronger light–matter coupling or enhanced UV absorption is desired (for example, UV detectors, photocatalysts, or absorber layers). Tensilely strained LAGO ($+z$ strain up to $+8\%$) becomes more transparent (lower n and k in the visible), shows weaker UV absorption, lower reflectivity, and lower-energy, weaker plasmonic response. This makes tensile-strained LAGO attractive for transparent dielectrics, antireflective coatings, and transparent electronics where minimal optical loss and low surface reflectance are needed. Because all optical quantities ($\varepsilon_1$, $\varepsilon_2$, n, k, $\alpha$, R, L) shift monotonically and predictably across the $-8 \rightarrow +8$ range, uniaxial z-strain provides an effective tuning handle to move critical spectral features (absorption edge, dielectric zero-crossing, plasmon peak) continuously across the UV window.

**Practical recommendations for characterization and device design:** Optical ellipsometry and UV-vis spectroscopy performed as a function of applied uniaxial strain (or by using coherently strained thin films on lattice-mismatched substrates) will directly reveal the predicted shifts in $\varepsilon_1$, $\varepsilon_2$, n, k, and $\alpha$. For device design, choose compressive strain states to maximize UV absorption/photocatalytic activity and tensile states when transparency/low reflectivity is paramount. Intermediate strains near 0% allow a balance between transparency and UV activity. When reporting results, present both the spectral shifts (photon-energy axis) and the relative change in peak intensity (oscillator strength) as functions of strain percentage; these two quantities together quantify how strain modifies both transition energy and probability.

Therefore, uniaxial strain along the c-axis from $-8\%$ to $+8\%$ tunes LAGO's optical response in a systematic way: compressive strain raises transition energies, enhances UV absorption and reflectivity, and hardens plasmon resonances, while tensile strain lowers transition energies, weakens absorption and reflectivity, and softens plasmonic response-making strain engineering a practical route to tailor LAGO for either active UV optoelectronic/photocatalytic roles or passive transparent/antireflective applications.

**Conclusions:**

In the present report, we conducted an inclusive first-principles investigation into the mechanical, electronic, and optical properties of $La_2AlGaO_6$ (LAGO) perovskite under applied uniaxial external strain. The calculated elastic constants satisfy the Born stability criteria, confirming the mechanical stability of LAGO. Analysis of the elastic constants revealed a greater resistance to linear compression along the c-axis, while a lower shear resistance was observed, indicating a susceptibility to deformation under shear stress. The positive tetragonal shear modulus and Kleinman parameter value of 0.55 highlight the dynamic stability of LAGO and validate that both bond bending and stretching contribute significantly to its mechanical strength. Bulk and shear moduli, evaluated via Voigt, Reuss, and Hill approximations, further support the material's mechanical robustness. The calculated Young's modulus and Poisson's ratio (0.26) suggest that LAGO is a stiff, brittle material, with significant ionic character in its chemical bonding. Pugh's ratio and Cauchy pressure analyses align with these findings, indicating a predominantly brittle nature and ionic bonding. Moreover, the machinability index of 1.77 points to excellent plasticity and dry lubrication potential, comparable to that of technologically important layered materials. Finally, the near-zero universal anisotropy factor confirms that LAGO is almost elastically isotropic. In addition, the strain effect on the electronic structure of LAGO perovskite has been comprehensively analyzed, revealing significant changes in the band gap and its nature. Based on the detailed analysis of strain-dependent electronic and optical properties of $La_2AlGaO_6$ perovskite, it is evident that uniaxial strain along the c-axis significantly modulates its band structure and optical response. Compressive strain enhances orbital hybridization between O-2p and Al/Ga p-states, leading to band gap narrowing and increased electronic delocalization, while tensile strain weakens this hybridization, resulting in band gap widening and more localized states. These variations are directly reflected in the optical spectra, where shifts in absorption edges, dielectric functions, and refractive indices correlate with the changes in electronic structure. Such tunability through strain engineering demonstrates the potential of $La_2AlGaO_6$ for advanced optoelectronic and strain-sensitive applications, where precise control over band gap and light-matter interactions is essential. The results highlight the sensitivity of the electronic structure to strain, which can be exploited to engineer materials with tailored properties for optoelectronic applications. These findings offer valuable insights into the potential of strain-engineering as a strategy for enhancing the performance of LAGO-based devices across various technological domains, including solar cells, light-emitting diodes, and piezoelectric devices. These results provide critical insights into the mechanical performance of $La_2AlGaO_6$, suggesting it as a promising candidate for applications requiring mechanical stability, thermal resilience, and good machinability in high-performance environments.

**Acknowledgments**

One of the authors, Debashis Bandyopadhyay, is thankful to the Science and Engineering Research Board (SERB), Government of India, for the research funding provided under sanction order CRG/2022/003249, which enabled this work.

## CRediT author statement

Miss Chaithanya Purushottam Bhat (first author) was primarily responsible for data analysis, organizing the research paper, conducting the literature survey, setting references, and reviewing and revising the manuscript draft. Miss. Joyti Dagar (2nd author) was responsible for the data correction, graphical draft checking, and optical property calculations. Mr. Ashwin K. Godbole (3rd author) was responsible for data analysis, writing the script, and checking the draft. Prof. Debashis Bandyopadhyay acted as the corresponding author and took on a broader set of duties, including project coordination, conceptualization, VASP simulations, manuscript writing, and review, as well as project administration. These CRediT statements offer a comprehensive and transparent overview of each author's contributions, ensuring proper acknowledgment of their roles in the project.

**Declaration of interests**

The authors confirm that they have no financial interests or personal relationships that could be perceived as conflicting with the research presented in this paper.

**Declaration of Generative AI and AI-assisted technologies in the writing process**

To improve the quality of the language and ensure proper citation, the author(s) used Grammarly and Turnitin software while drafting this manuscript. After utilizing these tools, the author(s) reviewed and made any necessary revisions, and they assume complete responsibility for the publication's content.

**Data availability declaration**

All data are provided in the manuscript, presented in table format.